# Distance dependent interaction between a single emitter and a single dielectric nanoparticle using DNA origami

Nicole Siegel[1§], María Sanz-Paz[1,2§*], Javier González-Colsa[3], Guillermo Serrera[3], Fangjia Zhu[1], Alan Szalai[4], Karol Kołątaj[1], Minoru Fujii[5], Hiroshi Sugimoto[5], Pablo Albella[3], and Guillermo P. Acuna[1,6*]

[1]Department of Physics, University of Fribourg, Chemin du Musée 3, Fribourg CH-1700, Switzerland.

[2]Sorbonne Université, CNRS, Institut des NanoSciences de Paris, INSP, F-75005 Paris, France.

[3]Group of Optics, Department of Applied Physics, University of Cantabria, 39005, Santander, Spain.

[4]Centro de Investigaciones en Bionanociencias (CIBION), Consejo Nacional de Investigaciones Científicas y Técnicas (CONICET), Godoy Cruz 2390, Ciudad Autónoma de Buenos Aires, C1425FQD, Argentina.

[5]Department of Electrical and Electronic Engineering, Graduate School of Engineering, Kobe University, Kobe 657-8501, Japan.

[6]Swiss National Center for Competence in Research (NCCR) Bio-inspired Materials, University of Fribourg, Chemin des Verdiers 4, CH-1700 Fribourg, Switzerland.

[§]equal contribution

Keywords: nanophotonics, dielectric nanoparticles, silicon, low-loss materials, single-molecule fluorescence, DNA nanotechnology, DNA origami.

**Abstract**

Optical nanoantennas can manipulate light-matter interactions at the nanoscale, modifying the emission properties of nearby single photon emitters. To date, most optical antennas are based on metallic nanostructures that exhibit unmatched performance in terms of electric field enhancement but suffer from substantial ohmic losses that limit their applications. To circumvent these limitations, there is a growing interest in alternative materials. In particular, high-refractive-index dielectrics have emerged as promising candidates, offering negligible ohmic losses, and supporting both electric and magnetic resonances in the visible and near-infrared range that can unlock novel effects. Currently, the few available studies on dielectric nanoantennas focus on ensemble measurements. Here, we exploit the DNA origami technique to study the interaction between silicon nanoparticles and organic fluorophores at the single molecule level, in controlled geometries and at different spectral ranges within the visible spectrum. We characterize their distance-dependent interaction in terms of fluorescence intensity and lifetime, revealing a significant modification of the decay rate together with minimal quenching and a high fluorescence quantum yield even at short distances from the dielectric nanoparticle. This work demonstrates the advantages of dielectric nanoantennas over their metallic counterparts and paves the way for their applications in single-molecule spectroscopy and sensing.

# 1. Introduction

Optical antennas mediate and enhance interactions between light and fluorescent emitters by modifying their local electromagnetic environment.[1][2] They can focus light beyond the diffraction limit by coupling free-space propagating light into highly enhanced and localized near-fields, thus increasing the excitation rate of nearby emitters.[3] Optical antennas can also tailor the emission properties of fluorescent sources through the manipulation of the local density of states (LDOS).[4][5] To date, the majority of the studies have focused on plasmonic optical antennas based on gold or silver nanostructures. With these metallic antennas, several effects were demonstrated, including the acceleration of the decay rates of fluorescent emitters[6] leading to giant luminescence enhancement,[3][7][8] ultrafast emission,[9][10] increased photostability,[11][12] and control of the emission pattern.[13]–[15] Despite all these advances, metals suffer from high ohmic losses throughout the visible range, which can lead to severe quenching of the fluorescence emission,[16]–[18] an effect that intensifies at short distances.[18]

Recently, high-refractive-index dielectric nanoparticles (HRID NPs), based on materials such as silicon, germanium or gallium phosphide, have been identified as promising alternatives to metals, since they can support Mie resonances in the optical range with weaker ohmic losses in the visible spectrum.[19][20] Although they exhibit more modest electric field enhancements than gold or silver optical antennas,[19][21] the absence of losses means that emitters in close proximity to HRID NPs will not suffer from quenching and could achieve a comparable overall fluorescence. In addition, a key difference with metallic nanostructures lies in the fact that HRID NPs can support both electric and magnetic resonances[21][22] which can give rise to a plethora of effects including directional scattering[23] and directional fluorescence emission,[24]–[26] the modification of the transition rates of electric and magnetic emitters[21][27]–[30] and the formation of super-chiral fields[31][32] for sensing applications.

To date, experiments on HRID optical antennas and their effect on nearby emitters have shown promising results in terms of fluorescence enhancement,[33]–[36] lower heating,[37] modification of decay rates[29] and reduced non-radiative quenching.[37][38] These results were obtained from ensemble measurements[35][37] or from diffusing molecules[33] where the distance to the antenna could not be controlled. Attempts to circumvent these limitations included the use of spacer layers[38] or scanning-probe microscopy,[29] however, with no stoichiometric control as an ensemble of emitters was addressed. In addition, it is worth mentioning that all these experiments were carried out with nanostructures obtained either by laser ablation or through the evaporation of the HRID materials. These synthesis approaches typically produce amorphous or polydisperse nanostructures with a larger imaginary part of their refractive index hindering its reproducibility.[39][40] Therefore, despite the great potential of HRID materials to advance the field of optical antennas and introduce new effects that have been extensively predicted,[20][24][41][42] an experimental study of the interaction between HRID NPs and fluorescent emitters at the single molecule level with nanometer precision and stoichiometric control is still missing.

Here, we use the DNA origami technique[43] to precisely place individual organic fluorophores and colloidal silicon nanoparticles[44] (SiNPs) at varying interdistances. These DNA-mediated constructs enable the study of the distance-dependent interaction between fluorophores and HRID NPs at the single-molecule level and in controlled geometries with high position accuracy. Furthermore, the SiNPs employed in this work are both crystalline and monodispersed and therefore exhibit a much lower imaginary part in the dielectric permittivity throughout the visible range compared to the amorphous Si nanostructures commonly used in lithography, potentially leading to lower non-radiative losses.[29][33] We carry out fluorescence intensity and lifetime measurements in two different spectral ranges that correspond to fractions of the spectrum where either the SiNP's electric or magnetic dipolar mode prevails.

Our results show for both spectral ranges that, while the radiative decay is enhanced close to the NP's surface, the non-radiative rate remains essentially unaltered at values smaller than the radiative decay rate, in contrast to the behavior observed for metallic nanoparticles.[18] This work characterizes the interaction between HRID NPs and organic fluorophores at the single-molecule level in highly controlled geometries and emphasizes the advantage of using dielectric materials for nanophotonic applications without the need to resort to different approaches to avoid quenching.[45][46] Finally, our results are supported by numerical simulations, which show good agreement with the experimental data, evidencing the outstanding positioning control enabled by the DNA origami technique for placing SiNPs.

2. Experimental results

**Figure 1A** depicts a schematic representation of the DNA origami employed. It consists of a two-layer 12-helix rectangular structure with dimensions of approximately 180 nm × 20 nm × 5 nm (length × width × height) and a "mast" in the center[47] (see further details about the DNA origami design in Tables S4 and S5). The DNA origami design includes 6 biotinylated modifications on the bottom side for their immobilization onto glass coverslips functionalized with neutravidin. Furthermore, the design also includes 32 binding sites consisting of a mixture of $A_8$ and $A_{18}$ single-stranded (ss) DNA handles to accommodate a single SiNP at the upper side (**Figure 1A**). Finally, a single organic fluorophore pair consisting of an ATTO 542 and an ATTO 647N, was incorporated. In total, three different DNA origami structures were prepared (**Figure 1B**), which differed only in the position of the fluorophore pair, covering an estimated distance to the SiNP's surface of approximately 7, 20 and 80 nm. The crystalline and monodispersed colloidal SiNPs[48] employed had a diameter of (140 ± 8) nm (see **Figure 1C** and details in the supporting information). For this size, SiNPs exhibit in water Mie resonances within the visible spectrum, see numerical simulations of the scattering cross section and mode decomposition in **Figure 1D**. Considering the two main contributions, the magnetic and electric dipolar modes with peaks at around 600 and 500 nm, respectively, throughout the emission range of the ATTO 542 the magnetic mode prevails, whereas for the ATTO 647N the situation is reversed (see **Figure 1D**).

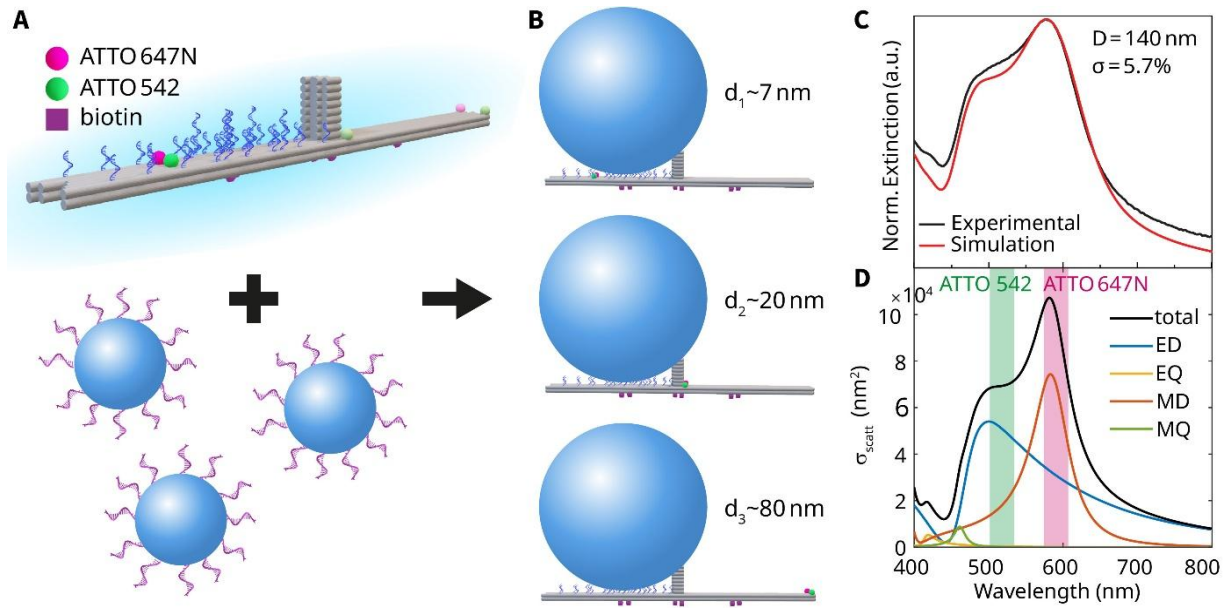

*Figure 1: Monomer assembly using DNA origami and SiNPs. (A) Sketch of the DNA origami used, displaying the different dye-NP distances (magenta and green dots) studied and the polyadenine (blue protrusions) strands to attach subsequently added DNA-functionalized SiNPs. Biotin molecules (violet diamonds) on the bottom of the origami are used for binding to the surface of a functionalized glass slide. (B) Illustration of the assembled monomer antennas for the three different dye-NP distances used ($d_1$, $d_2$ and $d_3$). (C) The extinction spectrum from a solution of SiNPs is measured (black) and compared to simulations (red). From these simulations, the average diameter $\bar{D}$ and the polydispersity $\sigma/\bar{D}$ of the NP distribution are obtained. (D) Scattering cross-section and mode decomposition for a SiNP of 70 nm radius in water. The shaded areas highlight the spectral emission bands of the two dyes used in our experiments (ATTO 542 in green and ATTO 647N in magenta).*

The incorporation of the SiNPs to the DNA origami structures was performed following a "surface synthesis" approach.[18] First, the DNA origami were immobilized onto glass coverslips through the biotin-neutravidin interaction. Second, a solution containing the previously DNA-functionalized SiNPs was added (see **Figure 1A**). After incubation, the solution containing the SiNPs was removed. We chose this approach over the "solution synthesis"[15] method since by adjusting the incubation time and NP concentration, the ratio of DNA origami hybridized with a single SiNP or none can be manipulated. This can be exploited to have a control reference population within the same sample that is being measured, which will be identified by subsequent colocalization with scanning electron microscopy (SEM). In addition, the surface synthesis approach yields a more consistent alignment of the SiNP and the fluorophore pair with respect to the incident light direction. In this way, dyes are located in the ("polar") plane below the SiNP, so that effects such as the enhancement of the excitation field can be neglected (see simulations of the electric field enhancement in **Figure S1**). To verify proper assembly and to estimate the incorporation yield, we first performed fluorescence lifetime imaging microscopy (FLIM) on the DNA origami sample with the fluorescent pair located at distance $d_1$ (**Figure 2A**). The results for the ATTO 647N dye show two distinct populations with lifetimes of around 4 ns, in agreement with the intrinsic lifetime of the dye, and shorter lifetimes of approximately 2.5 ns. These images were colocalized with the positions of the SiNPs as determined via SEM imaging (**Figure 2B**) of the same area. The overlay of both images (**Figure 2C**) reveals, first, no aggregation of SiNPs with minor non-specific binding of NPs to the surface. Second, the lifetime reduction can be ascribed to the incorporation of a single SiNP into the DNA origami, and with the incubation time and NP concentration employed, a significant fraction of the DNA origami contained no NP (for this case ~25 %). Therefore, spots colocalizing with a single SiNP were used for further analysis, and the ones without NP were taken as a reference.

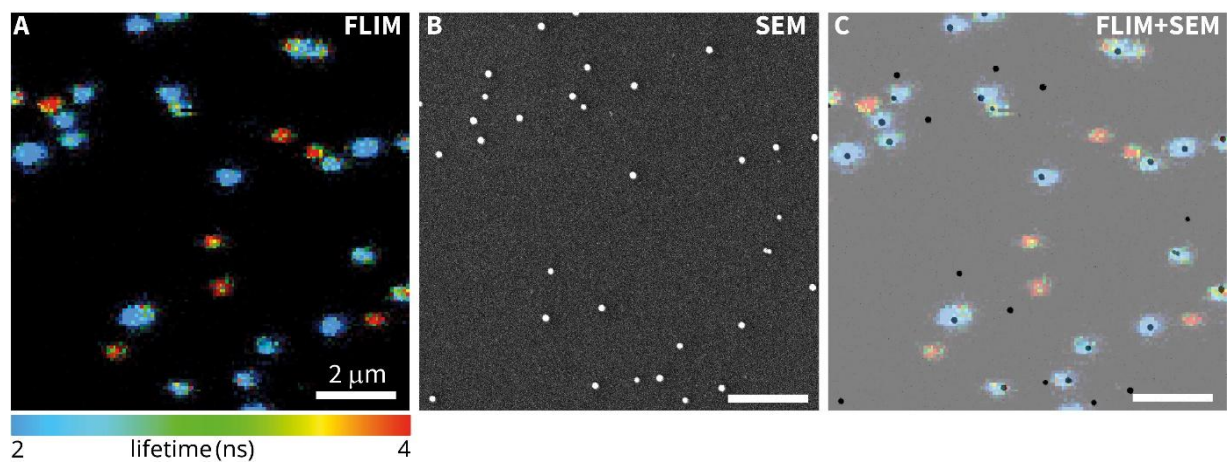

*Figure 2: Monomer imaging and colocalization. (A) Fluorescence lifetime image from ATTO 647N at $d_1$ and (B) corresponding SEM image of a region showing colocalization between fluorescent spots and SiNPs, (C) Overlay between both. Scale bars: 2 μm.*

To investigate the effect of a SiNP on the emission rates of a single dye, we employed fluorescence lifetime microscopy to extract the fluorescence intensity and lifetime of individual dye-SiNP constructs. The results for the ATTO 647N at distances $d_1$, $d_2$ and $d_3$ are included in **Figure 3A-C**, respectively, whereas the corresponding results for the ATTO 542 are included in **Figure 3D-F**. The reported fluorescence intensity is normalized to the average fluorescence intensity of the reference population, whereas for the fluorescence lifetime we used the absolute value since it is less prone to deviations due to alignment or slight variations in laser power. As expected, the measurements on the reference population yield a rather compact distribution with similar average fluorescence lifetimes for each dye. For the dye-SiNP hybrids at distance $d_3$, in the green and red spectral ranges, both the fluorescence intensity and lifetime overlap with the reference values. Therefore, we conclude that at distances > 80 nm the interaction is negligible. The situation is different for distances $d_1$ and $d_2$ in which a clear deviation from the reference values can be observed for both dyes, pointing to the broadband effect of SiNPs.
8

To better evaluate this distance-dependent trend, we plot in **Figure 4A,B** the median relative fluorescence lifetime $\frac{\tau}{\tau^o}$ and intensity $\frac{I}{I^o}$ enhancement for both dyes as a function of the distance to the NP surface. Here, the error bars represent the Median Absolute Deviation (MAD). The main features of **Figure 4** can be summarized in a lifetime reduction at $d_1$ and $d_2$ for both the ATTO 647N and the ATTO 542 fluorophores, stronger for the latter due to a higher spectral overlap, while at the same time, the fluorescence intensity remains essentially unchanged regardless of the dye or position.[38] These two observations hint at the conclusion that the radiative decay rate dominates the lifetime decrease,[49] even at this short distance, as opposed to what happens with gold or silver NPs.[50]

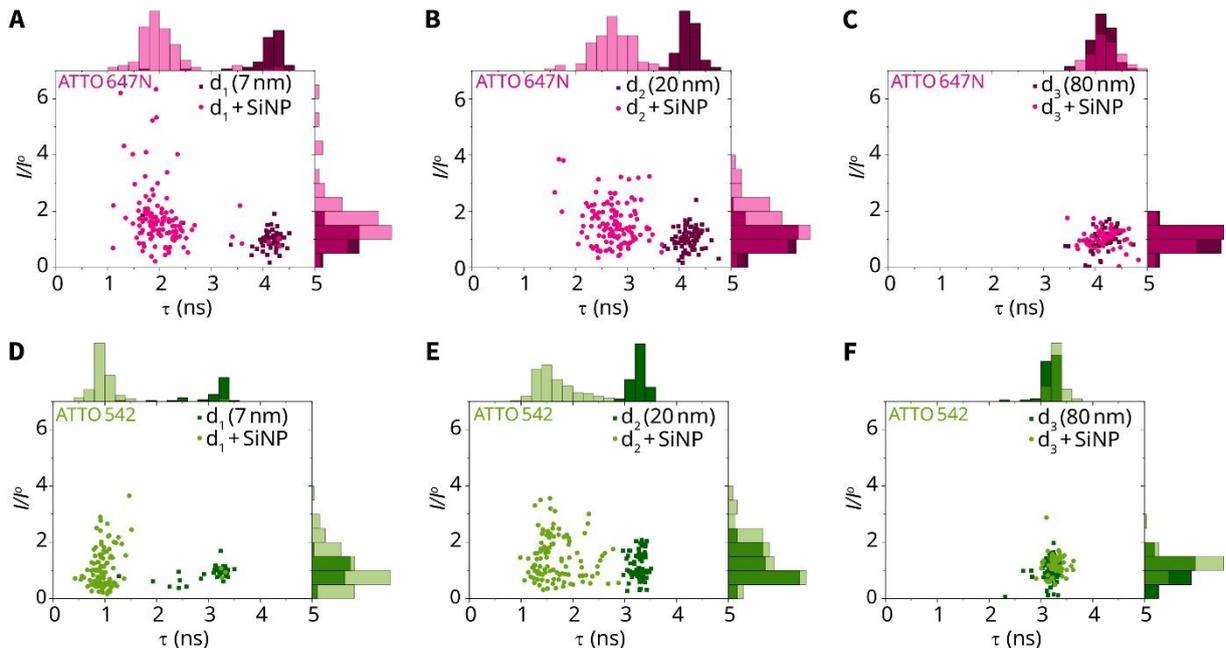

*Figure 3: Broadband lifetime reduction close to SiNPs. Intensity enhancement $I/I^o$ vs lifetime $\tau$ for the two dyes present in our DNA origami construct: (A, B, C) ATTO 647N and (D, E, F) ATTO 542, located at various distances ($d_1 \approx$ 7 nm, $d_2 \approx$ 20 nm, $d_3 \approx$ 80 nm) to the SiNP (lighter color) and in absence of the NP (darker color). Number of events measured (A) $N_{ref}$ = 55, $N_{Si}$ = 122, (B) $N_{ref}$ = 81, $N_{Si}$ = 126, (C) $N_{ref}$ = 71, $N_{Si}$ = 58, (D) $N_{ref}$ = 41, $N_{Si}$ = 112, (E) $N_{ref}$ = 74, $N_{Si}$ = 108, (F) $N_{ref}$ = 60, $N_{Si}$ = 59.*

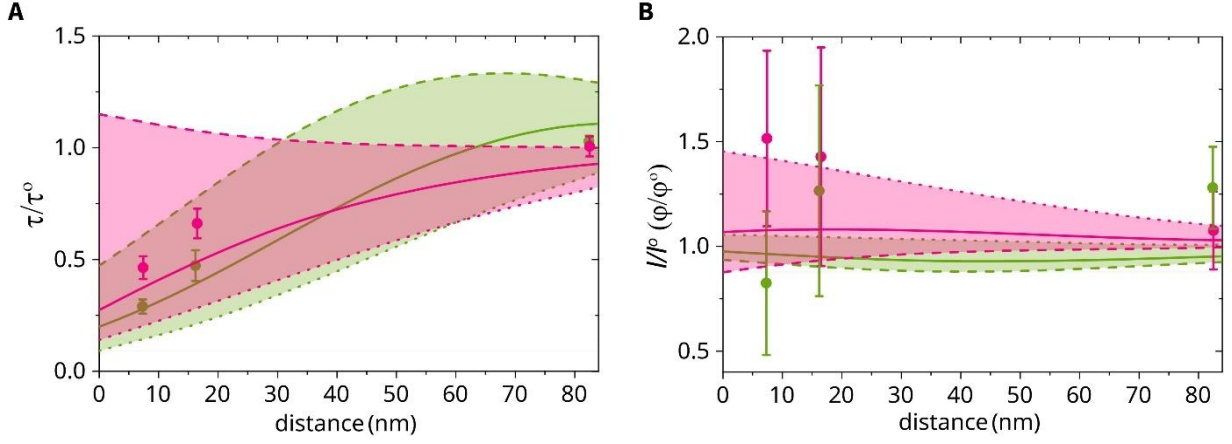

*Figure 4: Distance-dependent lifetime and intensity changes. Changes in lifetime τ (A) and intensity I (B) with respect to the isolated molecule as a function of the dye-SiNP distance for the two emitters: magenta for ATTO 647N and green for ATTO 542. Shadowed areas indicate the range between the simulated values for radial (dotted) and tangential (dashed) orientations, and the solid line represents the trend for an average orientation. Note that simulations in (B) correspond to changes in quantum yield φ. Experimentally obtained values (dots representing the median and bars the median absolute deviation) fall well within those expected ranges.*

To support this observation, we used those two experimentally measured magnitudes (fluorescence intensity and lifetime) to determine the distance-dependent modification of the decay rates for a single molecule in the vicinity of a NP and compared it to the isolated case.[17][50] First, the relative change in quantum yield $\frac{\varphi}{\varphi^o}$ can be approximated to the relative change in intensity $\frac{I}{I^o}$[17]

$$\frac{I}{I^o} = \frac{\varphi}{\varphi^o}\frac{k_{exc}}{k_{exc}^o}\frac{\eta}{\eta^o} \rightarrow \frac{\varphi}{\varphi^o} = \frac{I}{I^o}\frac{k_{exc}^o}{k_{exc}}\frac{\eta^o}{\eta} \approx \frac{I}{I^o} \qquad (1)$$

where the superscript $x^o$ corresponds to the reference case (isolated dye without NP), $k_{exc}$ refers to the excitation rate (which is proportional to the local electric field intensity), and $\eta$ is the collection efficiency. In **Equation 1**, we have neglected changes in the overall product of $\frac{k_{exc}^o}{k_{exc}} \frac{\eta^o}{\eta}$ as expected for the geometry of the hybrid system with the dye located in the polar plane. However, a thorough analysis to account for these effects can be found in **Figure S1** and **Figure S2**. Then, using this change in quantum yield together with the measured lifetimes ($\tau^o$ and $\tau$) and the intrinsic quantum yield $\varphi^o$ of the commercial dyes employed, the changes in decay rates can be obtained from:

$$\frac{k_r}{k_r^o} = \frac{\varphi}{\varphi^o} \frac{\tau^o}{\tau} \quad (2)$$

$$\frac{k_{nr}}{k_r^o} = \left(\frac{1}{\varphi^o} - \frac{\varphi}{\varphi^o}\right)\frac{\tau^o}{\tau} \quad (3)$$

with $k_r$ and $k_{nr}$ the radiative and non-radiative decay rates, respectively. These rates are estimated for both dyes at the three different distances, and the resulting values are shown in **Figure 5** (median and error bars in green for ATTO 542 and magenta for ATTO 647N). For the two different spectral ranges sampled, $k_r$ increases closer to the SiNP surface (**Figure 5A**), reaching enhancement values close to 3. In contrast, $k_{nr}$ remains essentially independent of the distance to the SiNP with lower enhancement values < 1 (**Figure 5B**). The uncertainty in the determination of $k_{nr}$ is higher than that of $k_r$ as expected from the subtraction in **Equation 3**. In conclusion, changes in $k_r$ dominate the lifetime reduction even at short distances, as they are comparable or even one magnitude larger than changes in $k_{nr}$ (**Figure 4D**). This is radically different from what has been measured for gold, where the non-radiative rate surpasses the radiative one by one order of magnitude.[50] Despite this difference, the lifetime is still reduced around a SiNP. For the shortest distance, the lifetime is lowered on average by 2.5-fold, a reduction that is larger than previously measured values around a single NP.[35][38]

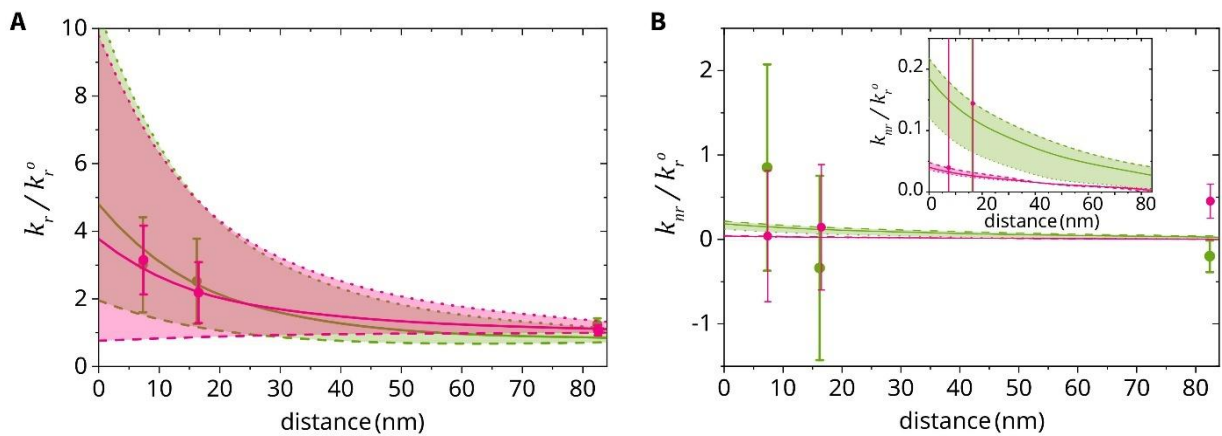

*Figure 5: Comparison between experimentally extracted rates and simulated values. Changes in radiative decay rate $k_r$ (A) and non-radiative decay rate $k_{nr}$ (B) with respect to the isolated molecule as a function of the dye-SiNP distance for the two emitters: magenta for ATTO 647N and green for ATTO 542. Shadowed areas indicate the range between the simulated values for radial (dotted) and tangential (dashed) orientations, and the solid line represents the trend for an average orientation. Experimentally obtained values (dots representing the median and bars the median absolute deviation) fall well within those expected ranges. Inset in (B) shows a zoom-in for better visualization of the simulated values.*

The experimental results are compared to numerical simulations performed for the two orientations of the fluorophore´s transition dipole moment (radial or tangential to the SiNP surface). These simulations are included in **Figure 4** and **Figure 5**, where dotted (dashed) lines represent the case of the radial (tangential) orientation, and the solid lines represent a weighted average (1 × radial + 2 × tangential). Since our measurements are carried out in an aqueous solution, dyes can be either freely rotating or fixed at any random orientation. Thus, the experimentally extracted rates are expected to fall in between those two extreme cases (shadowed areas in **Figure 4** and **Figure 5**). Indeed, we observe a good agreement with the simulations for the different parameters. The simulations also reveal, as opposed to the case of gold,[50] the low sensitivity of the non-radiative rate to the emitter orientation (**Figure 5B**).[51]

Note that changes in the collection efficiency or the excitation rate due to the SiNP itself have been neglected in **Figure 5**, i.e., $\frac{k_{exc}^o}{k_{exc}}\frac{\eta^o}{\eta} = 1$. A more detailed analysis taking into account those changes, has been included in **Figure S3** and shows that our conclusions remain unaltered with values that show good agreement with the simulations.

## 3. Conclusion

In conclusion, we have experimentally proven the effectiveness of silicon as an alternative to plasmonic materials for optical antennas, evidenced by the absence of fluorescence quenching of emitters at short distances to single SiNPs for two different spectral ranges within the visible regime. We have also extracted the changes in radiative and non-radiative decay rates from measured fluorescence intensity and lifetime values as a function of the distance to the SiNP. Our results show that changes in lifetime are dominated by an enhanced radiative decay rate, which can reach an order of magnitude more than changes in the non-radiative one. This opposes the established behavior for metals, where losses are the main source of decay at shorter distances to the NP. Such reduced quenching eliminates the need of spacer layers commonly used for sensing, opening promising possibilities for the use of high-refractive-index materials in the field. The results obtained are also rationalized through numerical simulations that exhibit a good agreement with the values of $\varphi/\varphi^0$, $\tau/\tau^0$, $\frac{k_r}{k_r^o}$ and $\frac{k_{nr}}{k_r^o}$ retrieved. This highlights the accuracy in distance control enabled by the DNA origami technique not only for metallic NPs, as it has been extensively shown, but also for high-refractive-index dielectric NPs. We believe that this work paves the way for more in-depth studies of light-matter interactions using dielectrics materials.

## 4. Experimental methods

*DNA functionalization of SiNPs.* Crystalline SiNPs were synthesized following established protocols[48] and size-separated using a sucrose gradient centrifugation process. Initially, a gradient forming system from Biocomp was employed to mix 50 wt % sucrose (placed at the bottom) and 20 wt % (placed at the top) using a SW40 rotor at 30 rpm, with a 12-second stop between rotations. The SiNP solution (0.5 wt % in methanol) was then carefully deposited on top, ensuring that the pipette tip made contact with the gradient surface to avoid breaking surface tension. The samples were centrifuged at 2000 rcf for 90 minutes. Size-separated SiNP solutions were retrieved using an automated piston that gradually descended from top to bottom at a rate of 0.5 mm sec$^{-1}$, collecting fractions every 3 mm. The solution was washed by centrifugation and resuspension twice with water and twice with methanol. Subsequently, SiNPs 0.04 wt % (diameter = 141 nm with SD~15 nm as calculated from the extinction simulated spectra, see supporting information) were dispersed in anhydrous DMF (10 mL) by ultrasonication, followed by the addition of CPTMS with a ratio of 200 CPTMS molecules per nm$^2$ of NPs. After 2 hours of sonication at 70°C, NaN$_3$ was added in excess, and the mixture was stirred overnight at 40°C. The NPs were then purified by centrifugation at 2000 rcf for 10 minutes, followed by consecutive redispersion twice in methanol and twice in water.[52] Finally, freezing-assisted SPAAC was employed for DNA conjugation. Shortly, SiNPs (1 mL, 0.1 wt %) were centrifuged at 2000 rcf for 10 minutes and resuspended in a 3:1 mixture of DBCO-modified T18 and T36 (300 μL, 100 μM, Biomers GmbH). Subsequently, 1 mL of PBS-SDS-Tween20 buffer (1X PBS pH 7.5, 0.1% SDS, 0.1 % Tween20) was added to the mixture, which was then frozen at -20°C for 2 hours. The NPs were thawed by sonication for 5 minutes, centrifuged twice for 10 minutes at 2000 rcf, and resuspended in water. For further purification and size separation, a 0.5 wt % agarose electrophoresis gel (LE Agarose, Biozym Scientific GmbH) in 0.5x TAE (20 mM Tris, 5 mM Acetate, 0.5 mM EDTA) and 6 mM MgCl$_2$ was run for 3 hours at 100 V. The NPs were recovered by extracting the desired band (green band corresponding to 140 nm) from the gel.[44] The extinction spectra of the NPs was measured, and the size was calculated (diameter = 140 nm with SD~8 nm as calculated from the extinction simulated spectra, see **Figure 1C**).

*DNA origami synthesis.* The DNA origami employed in this study was designed using CaDNAno,[53] and the structure files are available at https://nanobase.org/structure/146.[54] A 7249-nucleotide-long scaffold extracted from the M13mp18 bacteriophage (Bayou Biolabs LLC) was folded into the desired shape using 243 staples in 1xTAE (40 mM Tris, 10 mM Acetate, 1 mM EDTA), 12 mM $MgCl_2$, pH 8 buffer. It was mixed in a 10-fold excess of staples (purchased from IDT) over scaffold, and 100-fold for the functional staples (fluorophores, biotin, and handles, purchased from Biomers GmbH) shown in Tables S3 and S4. The mixture was heated to 70 °C and cooled down at a rate of 1 °C every 20 min up to 25 °C. The DNA origami structures were later purified by 1% agarose (LE Agarose, Biozym Scientific GmbH) gel electrophoresis at 70 V for 2 h and stored at 4 °C.[55]

*On-surface assembly of Optical Nanoantennas.* Glass slides with custom-made chromium grids were cleaned through a series of sonication baths, using the following solvents in sequence: 15 min in acetone, 15 min in isopropanol, 15 min in potassium hydroxide (3M), and 15 min in water. The surfaces were then activated by exposure to UV-ozone cleaning and subsequently coated with BSA-Biotin and Neutravidin (BSA-biotin and neutravidin, 0.5mg mL$^{-1}$ in PBS), each incubated on the grid for 25 min. The DNA origami (30 pM) was incubated on the surface for 15 min, and attached via Biotin protrusions from the bottom of the structure. Finally, 2 pM of functionalized SiNPs in PBS-SDS-Tween20 buffer (1X PBS pH 7.5, 0.1 % SDS, 0.1 %Tween20) were incubated overnight.

*Simulation of the decay rates.* The simulations included in **Figure 4** and **Figure 5** were performed using Lumerical FDTD together with the formalism described by Ringler et al.[56] Although fluorescence is a quantum mechanical effect, the radiative features of a fluorescent quantum emitter allow us to consider a classical approach. In this approximation, the emitter acts as a point-like dipole with a time-dependent dipole moment.[57] Therefore, we used a dipole source in two different contexts: a homogeneous medium with a water-like refractive index, and a non-homogeneous environment where the overall emission is affected by the presence of a dielectric nanoparticle immersed again in a water-like medium. Briefly, the enhancement factors of the frequency-dependent radiative $g_r$ and non-radiative $g_{nr}$ decay rates can be calculated from the far-field radiative $P_r$ and absorbed $P_{abs}$ powers,[57] respectively, and normalized to the radiated power $P_0$ of the dipole:

$$g_r(\omega) = \frac{P_r(\omega)}{P_0(\omega)} \quad (4)$$

$$g_{nr}(\omega) = \frac{P_{abs}(\omega)}{P_0(\omega)} \quad (5)$$

To calculate the radiative and absorbed powers, we use box monitors around the dipole, the nanoparticle, and the entire system. This allows us to independently integrate the power emitted by the dipole and the power absorbed by the nanoparticle, as well as to calculate the contribution of the whole system. It is important to note that with Lumerical FDTD the absolute decay rates cannot be calculated but only the relative ones. Furthermore, our computational model cannot calculate the intrinsic non-radiative decay rate since this decay includes non-radiative transitions coming from other phenomena, such as phonon excitations.

To ensure the adequacy of our simulations, we pay special attention to the mesh features, given the proposed dipole-particle distances. We select a minimum mesh step of 0.25 nm, which also allows for the correct reproduction of the nanoparticle curvature, preventing lightning rod effects.

As for the decay rate calculations, the total radiative $k_r$ and non-radiative $k_{nr}$ rates are given by the integrals

$$\frac{k_r}{k_r^o} = \int_0^\infty f_0(\omega) g_r(\omega) d\omega \tag{6}$$

$$\frac{k_{nr}}{k_r^o} = \int_0^\infty f_0(\omega) g_{nr}(\omega) d\omega \tag{7}$$

where $k_r^0$ is the radiative decay rate of the isolated molecule, and $f_0(\omega)$ is the integral-normalized fluorescence spectrum of the isolated dye:

$$f_0(\omega) = \frac{F_0(\omega)}{\int_0^\infty F_0(\omega) d\omega} \tag{8}$$

with $F_0(\omega)$ the fluorescence spectrum as provided in the vendor's data sheet for the two dyes used here.

From those quantities, the quantum yield can be calculated as

$$\varphi = \frac{\frac{k_r}{k_r^o}}{\frac{k_r}{k_r^o} + \frac{k_{nr}}{k_r^o} + \left(\frac{1}{\varphi^o} - 1\right)} \tag{9}$$

with $\varphi^o$ being the intrinsic quantum yield of the isolated molecule (0.65 for ATTO 647N and 0.93 for ATTO 542).

*Experimental setup.* Measurements were performed on an inverted microscope (Olympus IX71). Excitation was performed with a randomly polarized supercontinuum white light laser (FYLA SCT1000) that was spectrally filtered to a wavelength of (635 ±5) nm or (532 ± 5) nm to efficiently excite the dye (ATTO 647N or ATTO 542, respectively). A high NA objective was used (Olympus, 100× NA=1.4) for excitation and collection. Emitted fluorescence was spectrally split into two paths by a dichroic mirror (640DCXR, Chroma) and detected by two avalanche photodiodes (tau-SPAD, PicoQuant) with appropriate filters. The APDs and the pulsed laser are connected to a module for time-correlated single-photon counting (MultiHarp150, PicoQuant) for recording the photon arrival times. Measurements were performed as follows. First, a confocal image of a region of the sample was recorded, and the position of the DNA origami structures was determined. For each region imaged, measurements were taken first using red excitation to ensure all ATTO 647N dyes are bleached before moving to green excitation in order to avoid FRET between both fluorophores. Then, each structure was brought into the center of the confocal observation volume, and fluorescence transients were recorded. Only transients with single-step photobleaching were considered to guarantee that only single molecules were studied. From those transients, the average intensity before bleaching was extracted, together with the photon arrival times. Based on the latter, a decay histogram is obtained and fitted with a mono-exponential decaying function convoluted with the instrument response function (IRF) of the system to extract the fluorescence lifetime.

**Supporting Information**

**Calculations of silicon NP size distribution.** We estimate the dispersity of our SiNP solution using the measured extinction spectrum. We then tried to reproduce the measured extinction spectrum from the calculated spectra of single SiNPs by assuming a normal size distribution with average diameter $\bar{D}$ and standard deviation $\sigma$ as fitting parameters.[48] The measured and calculated spectra are shown in **Figure 1C**. The best fit obtained shows good agreement with the values obtained from TEM measurements, indicating high shape uniformity and dispersion. Under the Beer-Lambert law (i.e., no multiple scattering), extinction can be described as

$$Extiction = \varepsilon l c \sim 0.434\, \sigma_{ext}\, lN$$

where $\varepsilon$ is the extinction coefficient, $l$ the path length, $c$ the concentration, and $N$ the number of NPs. In a solution with a size distribution, the extinction cross-section $\sigma_{ext}$ can be written as a function of the NP's diameter $D$ and the probability density function $P$ as

$$\sigma_{ext} = \int_0^\infty \sigma_e(D,\lambda) P(D) dD$$

where $\sigma_e$ is the size and wavelength-dependent extinction cross section, obtained from simulations and used to compute the simulated extinction spectrum.

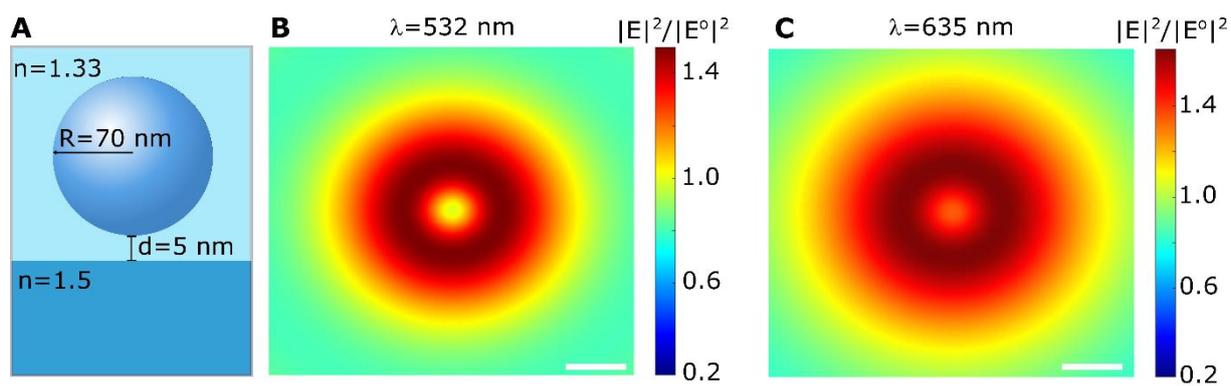

*Figure S1: Electric field enhancement at the emitter position plane.* Maps of the enhancement of the intensity of the electric field at the transversal middle plane between a SiNP (radius = 70 nm) and a glass substrate (A) at the two excitation wavelengths: 532 nm (B) and 635 nm (C). The values at the different positions where the emitter is placed are summarized in Table S1. Scale bar: 50 nm.

|  | $|E|^2/|E^o|^2$ @ $\lambda = 532$ nm | $|E|^2/|E^o|^2$ @ $\lambda = 635$ nm |
| --- | --- | --- |
| $d_1$ | 1.38 | 1.56 |
| $d_2$ | 1.47 | 1.63 |
| $d_3$ | 0.88 | 1.08 |

*Table S1. Electric field enhancement at the various emitter distances.* Extracted values from the maps in Figure S1 at the different emitter distances tested here.

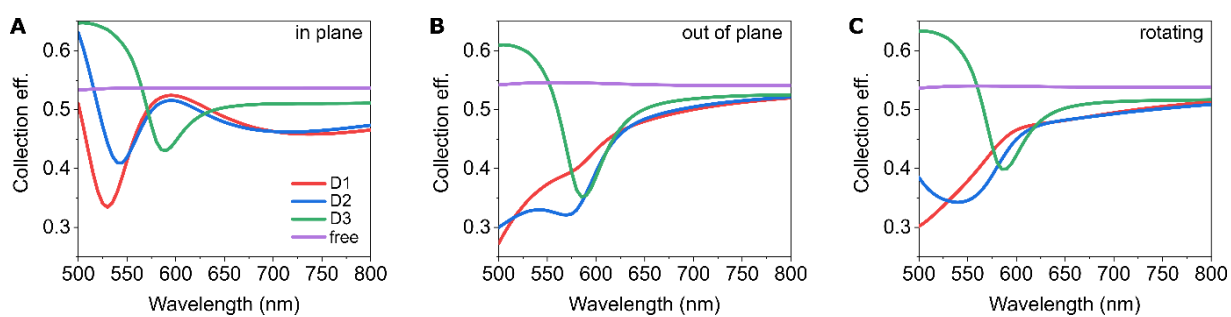

*Figure S2: Collection efficiency for an emitter at different distances to a SiNP.* *Collection efficiency at the glass side for a dipole sitting at a glass-water interface with in plane (A), out of plane (B) or averaged (rotating) (C) orientation placed at different distances ($d_1$, $d_2$ and $d_3$) from a SiNP (radius = 70 nm) as a function of its emission wavelength. The case of a dipole without any SiNP (free) is used as a reference.*

|        | $\eta/\eta^o$ for ATTO 542 | $\eta/\eta^o$ for ATTO 647N |
|--------|---------------------------|-----------------------------|
| $d_1$  | 0.85                      | 0.91                        |
| $d_2$  | 0.82                      | 0.91                        |
| $d_3$  | 0.93                      | 0.94                        |

*Table S2. Changes in collection efficiency at the various emitter distances.* *Displayed values are obtained by multiplying the values shown in Figure S2 with the normalized emission spectrum of each dye and integrating over the whole spectral range.*

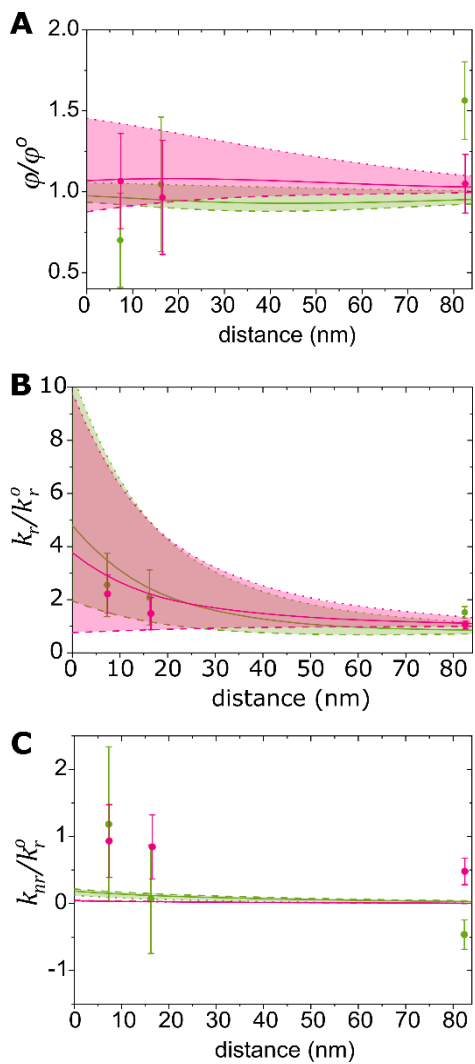

*Figure S3: Comparison between experimentally extracted rates (after applying a correction factor) and simulated values.* Changes in quantum yield $\varphi$ (A), radiative decay rate $k_r$ (B) and non-radiative decay rate $k_{nr}$ (C) with respect to the isolated molecule as a function of the dye-SiNP distance for the two emitters: magenta for ATTO647N and green for ATTO542. Shadowed areas indicate the range between the simulated values for radial (dotted) and tangential (dashed) orientations, and the solid line represents the trend for an average orientation. Experimentally obtained values (bars showing standard error around the median) fall well within those expected ranges. Intensity ratios obtained experimentally are converted into quantum yield ratios by multiplying by a correction factor that takes into account simulated changes in excitation rate and collection efficiency in the presence of the SiNP.

**List of DNA staples:**

| Name | Sequence |
|---|---|
| Biotin1 | ACAGGAAGATTGTCCCCCTTATTCACCCTCATTTGTTTC-Biotin |
| Biotin2 | GTTGATAGATATAAGCATAAGTATAGC-Biotin |
| Biotin3 | AGAGTACTCACGCTAACCTTTAATTGC-Biotin |
| Biotin4 | CACTAAAACACTCACGAACTAACACTAAAGT-Biotin |
| Biotin5 | TCACGACGTTGGGCGCTTTGGTAAAAC-Biotin |
| Biotin6 | CAGAGATAGCGATAGTGAATAACATAA-Biotin |
| ATTO542_P1 | AATTGAGTAATATCAGAAAATAAACAGCCATA-ATTO542 |
| ATTO647N_P1 | TACCCCGGTTAAAATTCCTTTGCCCGAACGTT-ATTO542 |
| ATTO542_P2 | ATTO542-CCCCCGCTGAGAGCCAGCAGGCCTGCAACAGTGCCACCCC |
| ATTO647N_P2 | ATTATCACCGGAAATGTTAGCAAACGTAGAA-ATTO647N |
| ATTO542_P3 | ATTATTACTTGGGAAGTTCATTACCCAAATCA-ATTO647N |
| ATTO647N_P3 | ATTO647N-CCCCCAGAGCGGGAGCTAACTTTCCTCGTTAGAATCCCC |

*Table S3:* *List of biotinylated and dye-modified staple sequences used in this project compared to the original design.* [47] *ATTO 647N and ATTO 542 staples present a single ATTO 647N or ATTO 542 fluorescent molecule for the alternative positions (P1, P2 and P3), and Biotin staples allow binding to a neutravidin-functionalized glass surface.*

| Name | Sequence |
|---|---|
| H.left 1(A8) | TACAAATTGCCAGTAAAGTAATTCTGTCCAGAAAAAAAA |
| H.left 2(A8) | TAAAGACTGTTACTTAGGCGCAGACGGTCAATAAAAAAAA |
| H.left 3(A8) | GCCACCCTTCGATAGCATAATCCTGATTGTTTAAAAAAAA |
| H.left 4(A8) | CACCAACCAAGTACAAGTACAGACCAGGCGCAAAAAAAA |
| H.left 5(A8) | CGACATTCCCAGCAAAATTATTTGCACGTAAAAAAAAAAA |
| H.left 6(A8) | CTTTTTTTTCATTTCAACAATAACGGATTCGAAAAAAAA |
| H.left 7(A8) | ACGGGTAATAAATTGTTGACCAACTTTGAAAGAAAAAAAA |
| H.left 8(A8) | ATTAATTATGAAACAATATACAGTAACAGTACAAAAAAAA |
| H.left 9(A8) | AATATTGACGTCACCGTGCGTAGATTTTCAGGAAAAAAAA |
| H.left 10(A8) | AACAGTAGCCAACATGACATGTTCAGCTAATGAAAAAAAA |
| H.left 11(A8) | AAGAACGCAAGCAAGCATAATATCCCATCCTAAAAAAAAA |
| H.left 12(A8) | CTTGCGGGGTATTAAAAAACCAATCAATAATCAAAAAAAA |
| H.left 13(A8) | AAATCAGATCATTACCATCAACAATAGATAAGAAAAAAAA |
| H.left 14(A8) | CATATGGTAGCAAGGTAATGGAAGGGTTAGAAAAAAAAA |
| H.left 15(A8) | CCATCTTTCGTTTTCAAACCACCAGAAGGAGCAAAAAAAA |
| H.left 16(A8) | GAGCCACCATCAAGTTTCCTGATTATCAGATGAAAAAAAA |
| H.left 17 (A18) | AAAAAAAAAAAAAAAAAAGGATTATACTTCTGAACCGGAAACGTCACCAA |
| H.left 18 (A18) | AAAAAAAAAAAAAAAAAAACCTACCATATCAAAATCACCAGTAGCACCA |
| H.left 19 (A18) | AAAAAAAAAAAAAAAAAAATTTACGAGCATGTAGCCAAGTACCGCACTCA |
| H.left 20 (A18) | AAAAAAAAAAAAAAAAAACATAAGGGAACCGAACGTCGAAATCCGCGACC |
| New Core 1 | GAAACGCAAAGACACCGCCAAAGATACCGAAG |
| New Core 2 | ATTTTGCACCCAGCTATTAGCGAAAGAATTAA |
| New Core 3 | AGCGAAAGACAGCATCAGGAAGTTTGTAGCAT |
| H.left 22(A18) | AAAAAAAAAAAAAAAAAAGCCGCCACTCATCAATAGCACCGTAATCAGT |
| H.left 23(A18) | AAAAAAAAAAAAAAAAAAATGGCAATCAGAACCCGCCTCCCTCAGGAGG |
| H.left 28(A8) | TGAGCGCTTAAGCCCATGGCATGATTAAGACTAAAAAAAA |
| H.left 29(A8) | CTGAACACATAGCAATAAACGCAATAATAACGAAAAAAAA |
| H.left 30(A8) | AAACAGGGAAGAAAAGGCCGAACAAAGTTACAAAAAAAA |
| H.left 31(A8) | TCCACAGAGGTGTATCGGATAAGTGCCGTCGAAAAAAAA |
| H.left 32(A8) | GTCACCAGCGCCACCCTTAGGATTAGCGGGGTAAAAAAAA |
| H.left 33(A8) | TAGGAACCCCACCCTCTATTAAGAGGCTGAGAAAAAAAAA |
| H.left 34(A8) | AATTAGAGAACCGATTGGCAACATATAAAAAAAAAAAAA |
| H.left 35(A8) | TTACCATTTTACCAGCACGGAATAAGTTTATTTAAAAAAAA |
| H.left 36(A8) | TGAAACCACAGAACCGAGCCACCACCCTCAGAAAAAAAAA |
| H.left 37(A8) | GCGACAGAACCGGAACACCACCAGAGCCGCCGAAAAAAAA |

*Table S4:* *List of DNA modified staples, which were extended as NP capturing strands, used for the directional antennas compared to the original origami design.* [47] *Handles are staples extended with an PolyA (number of A indicated in parenthesis) to bind a single PolyT functionalized SiNP.*